\documentclass[journal=jacsat,manuscript=article]{achemso}

\usepackage{xcolor}
\usepackage{lineno}
\usepackage[version=3]{mhchem}
\usepackage{booktabs}
\usepackage{amssymb}

\usepackage{pdfpages}

\author{Chetana Badala Viswanatha}
\affiliation{Prayoga Institute of Education Research, Bengaluru 560116, India}
\alsoaffiliation{Mindbrug Research LLP, Bengaluru 560085, India}
\email{chetana@mindbrug.com}
\author{Ka Man Yu}
\affiliation{Department of Physics and Research Center OPTIMAS, RPTU University Kaiserslautern-Landau,
Erwin-Schr\"{o}dinger-Stra{\ss}e 46, 67663 Kaiserslautern, Germany}
\author{Benito Arnoldi}
\affiliation{Department of Physics and Research Center OPTIMAS, RPTU University Kaiserslautern-Landau,
Erwin-Schr\"{o}dinger-Stra{\ss}e 46, 67663 Kaiserslautern, Germany}
\author{Anagha Aravind}
\affiliation{Prayoga Institute of Education Research, Bengaluru 560116, India}
\author{Aaruni Kaushik}
\affiliation{Department of Mathematics, RPTU University Kaiserslautern-Landau,
Erwin-Schr\"{o}dinger-Stra{\ss}e 46, 67663 Kaiserslautern, Germany}
\author{Jannis Lessmeister}
\affiliation{Department of Physics and Research Center OPTIMAS, RPTU University Kaiserslautern-Landau,
Erwin-Schr\"{o}dinger-Stra{\ss}e 46, 67663 Kaiserslautern, Germany}
\author{Martin Aeschlimann}
\affiliation{Department of Physics and Research Center OPTIMAS, RPTU University Kaiserslautern-Landau,
Erwin-Schr\"{o}dinger-Stra{\ss}e 46, 67663 Kaiserslautern, Germany}
\author{Benjamin Stadtmüller}
\affiliation{Experimentalphysik II, Institute of Physics, University of Augsburg, 86159 Augsburg, Germany}
\author{S. Harshini Tekur}
\affiliation{Prayoga Institute of Education Research, Bengaluru 560116, India}
\alsoaffiliation{Mindbrug Research LLP, Bengaluru 560085, India}
\email{harshini@mindbrug.com}

\title[Chirality from Atomic Models and Fermi Surfaces]
{Decoding Crystallographic Surface Chirality from Real Space and Momentum Space Images by Machine Learning}
\abbreviations{ARPES, CNN, CIP, fcc, TTA, STM}

\begin{document}

\begin{tocentry}
\includegraphics[width=\textwidth]{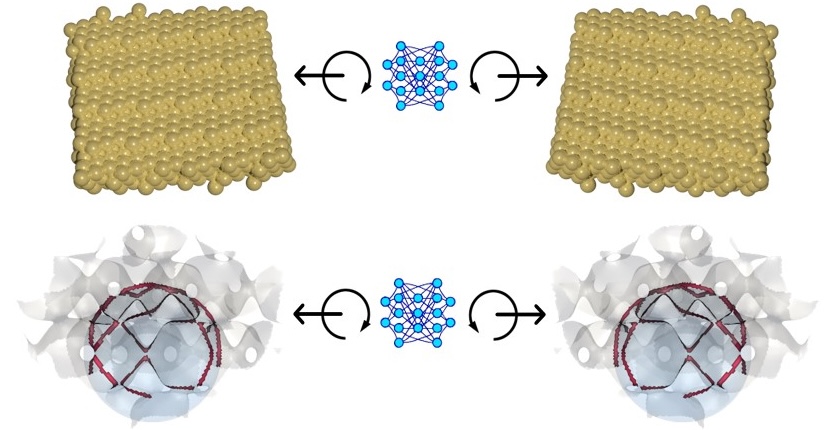}
\end{tocentry}

%%%%%%%%%%%%%%%%%%%%%%%%%%%%%%%%%%%%%%%%%%%%%%%%%%%%%%%%%%%%%%%%%%%%%
%% Abstract
%% CHANGE: removed cross-domain correlation sentence
%%%%%%%%%%%%%%%%%%%%%%%%%%%%%%%%%%%%%%%%%%%%%%%%%%%%%%%%%%%%%%%%%%%%%
\begin{abstract}
Intrinsically chiral metal surfaces, where handedness arises from the
asymmetric step-kink-terrace topology of high-Miller index planes, are
model systems for enantiospecific catalysis, sensing, and spintronics.
Yet no consistent method exists to classify their handedness directly
from experimental observables, without prior knowledge of crystallographic indices. Here, we report a dual-domain machine
learning framework that recognizes handedness in chiral metal surfaces from two independent image representations: atomic structure models in
real space and simulated momentum-resolved photoemission maps of Fermi
surface projections in reciprocal space. A ResNet18 model pretrained on general image data is fine-tuned for chirality classification independently on both image representations and achieves meaningful classification accuracy for both real space and reciprocal space images. However, the Fermi surface classifier significantly outperforms the real space classifier. Critically, the reciprocal space classifier, trained only on synthetic images, correctly identifies synchrotron-acquired experimental Angle-Resolved Photoemission Spectroscopy (ARPES) maps of Cu(643)$^R$ and Cu(643)$^S$. The stronger performance in momentum space, as well as the transfer to experimental data, shows that handedness
is encoded more globally and robustly in the electronic structure than
in the spatially localized kink-site geometry. These results establish
ARPES as a quantitative readout of crystallographic surface chirality
and suggest a scalable route to identify chiral metal surfaces relevant
to spin-selective phenomena.
\end{abstract}
%%%%%%%%%%%%%%%%%%%%%%%%%%%%%%%%%%%%%%%%%%%%%%%%%%%%%%%%%%%%%%%%%%%%%
\section{Introduction}
%%%%%%%%%%%%%%%%%%%%%%%%%%%%%%%%%%%%%%%%%%%%%%%%%%%%%%%%%%%%%%%%%%%%%
Chirality is a fundamental symmetry property that governs stereoselectivity
in catalysis, and has applications in sensing and spin-polarized transport.
Its most striking manifestation in condensed matter is the chiral-induced
spin selectivity (CISS) effect, which links structural handedness to
spin-polarized charge transport.\cite{cissname, naaman2019,
spincontrolchemistry}
CISS has been extensively studied in chiral organic molecules and inorganic
solids,\cite{helicenemunster, ciss2023outlook1} where spin-orbit coupling,
geometric chirality, and dephasing act cooperatively to produce spin
selectivity.\cite{2026_CISSdevicePRBL}
Crucially, spin selectivity vanishes immediately in achiral configurations,
identifying geometric chirality as the operative symmetry-breaking
ingredient.\cite{2026_CISSdevicePRBL}
This has direct implications for spin-active devices, spin-dependent
electrocatalysis, and enantioselective electrochemistry.\cite{2026_CISSdevicePRL,
tripletoxygenciss, chiralelectrochem2024}
Yet its manifestation at intrinsically chiral metal surfaces, where the
chirality is entirely geometric rather than molecular, remains largely
unexplored.

Even in highly symmetric face-centered cubic (fcc) metals such as Cu and Pt, chirality can arise at distinct surfaces from the mere geometric arrangement of surface atoms without any chemical diversity.
High-Miller index planes of the form $(h\,k\,l)$, where $h \times k \times l \neq 0$ and
$h \neq k \neq l$, expose kink sites whose asymmetric
step--kink--terrace topology breaks mirror symmetry, yielding intrinsically
chiral surfaces classified as R or S.\cite{crysnomattard1, crystalnomenclature,
crystalnomenclature2}
These surfaces are established model systems for enantiospecific heterogeneous
catalysis.\cite{gellman2021, Huang_Li_Duan_Li_Zhang_Dong_Wu_Huang_Zhang_Ding_etal_2026}
Cu(643) is an exemplary case: its bare geometric chirality is sufficient
to reverse spin selectivity in an all-chiral metal--molecule heterostructure,
directly linking surface kink geometry to CISS\cite{viswanatha2022}. This establishes intrinsically chiral metal surfaces as
a compelling and underexplored platform for spin-based applications.
Realizing this potential requires reliable methods to identify and classify
surface handedness across the full family of chiral metal surfaces.

The R/S handedness convention is defined in real space through atomic models
of the kink site geometry.
Whether this handedness extends directly into momentum space, and whether
it can be decoded from the electronic structure alone, is an open question.
Angle-resolved photoemission spectroscopy (ARPES) maps constant-energy cuts
through the electronic structure, yielding Fermi surface projections that
carry the broken mirror symmetry of the chiral surface in reciprocal
space.\cite{hufnerARPES, ARPES2019, 2026_CISSdevicePRL}
Here, we demonstrate that the R/S handedness is directly preserved in ARPES
Fermi surface maps: the clockwise or anticlockwise sense that defines R and S
in real space is reflected in the rotational orientation of the corresponding
Fermi surface polygons in reciprocal space.
The space of symmetry-inequivalent chiral $(h\,k\,l)$ planes is vast, yet
only a small subset has been characterized experimentally to date, and
experimental ARPES data on intrinsically chiral metal surfaces are
particularly rare. Therefore, making a simulation-based training strategy is the only
practical route to systematically classify surfaces as R or S across this broad Miller index family.

Machine learning (ML) has begun to address analogous classification
problems in crystallography\cite{Ziletti_Kumar_Scheffler_Ghiringhelli_2018, Leitherer_Ziletti_Ghiringhelli_2021}, with deep learning being deployed for classification of crystal space
groups from electron diffraction patterns\cite{Aguiar_Gong_Unocic_Tasdizen_Miller_2019}, and surface types from microscopy images\cite{Li_Yin_Yang_Du_Lu_2026, Burzawa_Liu_Carlson_2019, Gui_Zhang_Li_Luo_Xia_Wu_Chu_2023, Sokolov_2024}. Studies on machine learning applied to identification and classification of chirality have mostly been restricted to chiral molecules\cite{Zhou_Zhu_Yuan_Song_Mort_2025, Li_Telychko_Yin_Zhu_Li_Song_Yang_Li_Wu_Lu, Peng_Yu_Shi_Chen_Wang_Du_Huo_Yang_2025, Seifert_Stritzke_Kasten_Moller_Fingscheidt_Etzkorn_De_Wolff_Schlickum_2024}, or materials\cite{Groschner_Pattison_Ben-Moshe_Alivisatos_Theis_Scott_2022, Visheratina_Visheratin_Kumar_Veksler_Kotov_2023, Huang_Chu_Wu_Niu_Yu_Ma_2025} with large datasets. Prior applications have typically relied on backbones pretrained on domain-specific chemical or crystallographic databases~\cite{Pan_Yang_2010, Kaufmann_Lane_Liu_Vecchio_2021}. Meanwhile, transfer learning on pretrained architectures like ResNet\cite{ResNet_He_Zhang_Ren_Sun_2016} has proven effective when
domain-specific training data are scarce\cite{Pan_Yang_2010, Kaufmann_Lane_Liu_Vecchio_2021, Oviedo_Ren_Sun_Settens_Liu_Hartono_Ramasamy_DeCost_Tian_Romano_etal_2019}, as deep neural networks require large amounts of labelled data to train reliably from scratch. Transfer learning circumvents this by starting from a network already trained on a vast general-purpose image dataset, so that the model has already learned broadly useful visual features such as edges, textures and shapes. Here, we fine-tune ResNet18\cite{ResNet_He_Zhang_Ren_Sun_2016} (a deep learning architecture used primarily for image recognition and classification tasks) on two independent image representations: real space atomic models of high-Miller index metal surfaces, and simulated ARPES Fermi surface projections in reciprocal space. We use a purely vision-pretrained backbone with no chemical prior, demonstrating that generic visual features are sufficient to learn the symmetry-breaking signature of crystallographic surface chirality. Vitally, the reciprocal space classifier transfers directly to synchrotron-acquired
experimental ARPES maps of Cu(643)$^R$ and Cu(643)$^S$, correctly identifying both enantiomers. This constitutes our main result: we establish that ARPES Fermi surface maps capture the rotational asymmetry signature for chiral metal surfaces and globally encode geometric surface chirality in electronic momentum space. The ML model developed here, trained only on simulated Fermi surface maps, transfers successfully to experimental data across a substantial domain shift. And because the chirality signature arises from step–kink–terrace topology rather than chemical identity, we expect the same strategy to extend to other fcc metals and, with minimal retraining, to body-centred cubic (bcc) surfaces.\cite{bccFe}

\section{Results and Discussion}

\subsection{Physical Basis for Chirality in Momentum Space}

\begin{figure}
\centering
%% PLACEHOLDER: Figure 3
%% Two panels:
%% (a) Experimental ARPES map for the R surface, annotated with
%%     classifier-predicted label ("R") and sigmoid confidence score.
%% (b) Experimental ARPES map for the S surface, annotated with
%%     classifier-predicted label ("S") and sigmoid confidence score.
%% Each panel also shows the surface normal marker (filled circle)
%%     used as the geometric reference for the classifier.
\includegraphics[width=\textwidth]{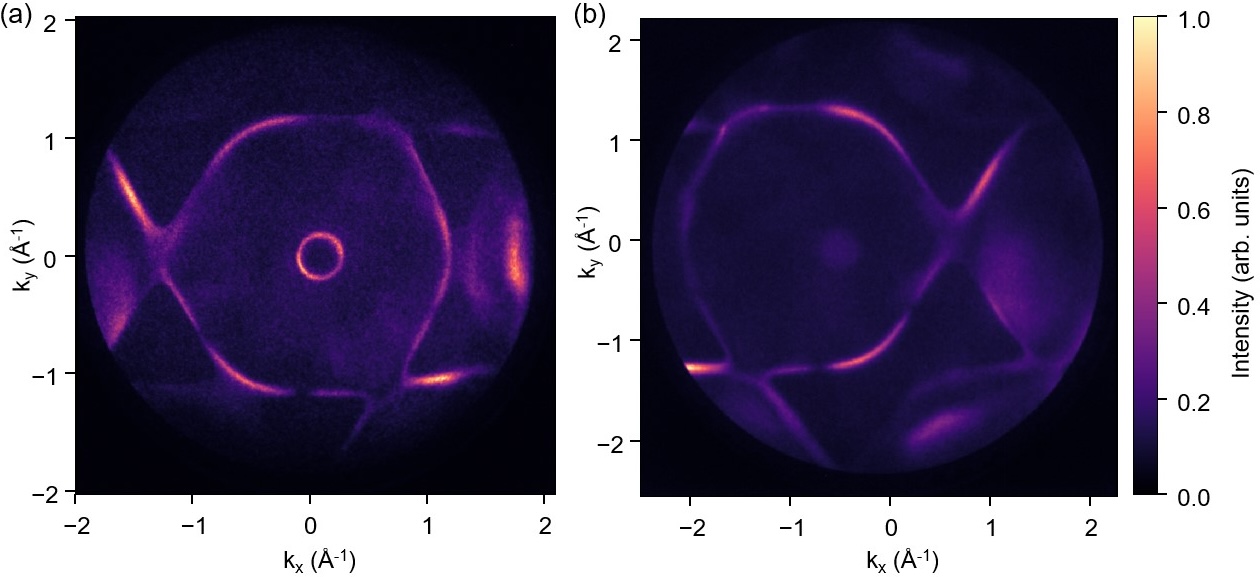}
\caption{Experimentally acquired Fermi surface maps for the (a) Cu(111) and (b) Cu(643)
surfaces, respectively (h$\nu=30\,$eV).}
\label{fig:arpes_chiral}
\end{figure}

We begin by establishing the physical basis for chirality in the momentum space representation. The momentum-resolved photoemission data in Figure~\ref{fig:arpes_chiral} constitute, to our knowledge, the first experimental demonstration that intrinsically chiral surfaces possess a correspondingly chiral Fermi surface. The Fermi surfaces of Cu(111) and Cu(643), shown side by side in Figure~\ref{fig:arpes_chiral}, illustrate this directly: the achiral Cu(111) surface yields a mirror symmetric Fermi surface, whereas the chiral Cu(643) surface does not. Early identification of chiral surfaces relied on similar mirror symmetry breaking of two-dimensional diffraction patterns such as
low-energy electron diffraction (LEED).\cite{crystalnomenclature} 
% \textcolor{red!60!black}{This correspondence raises a natural question: can the R/S classification convention, originally defined for real-space atomic arrangements, be generalized across representations? We demonstrate here through our datasets that ARPES provides a natural momentum space counterpart to this convention, as it maps the electronic structure onto a two-dimensional reciprocal-space image that can be compared directly with the synthetic Fermi surface projections.}

\subsection{Datasets and Workflow}

\begin{figure}
\centering
\includegraphics[height=10.5cm]{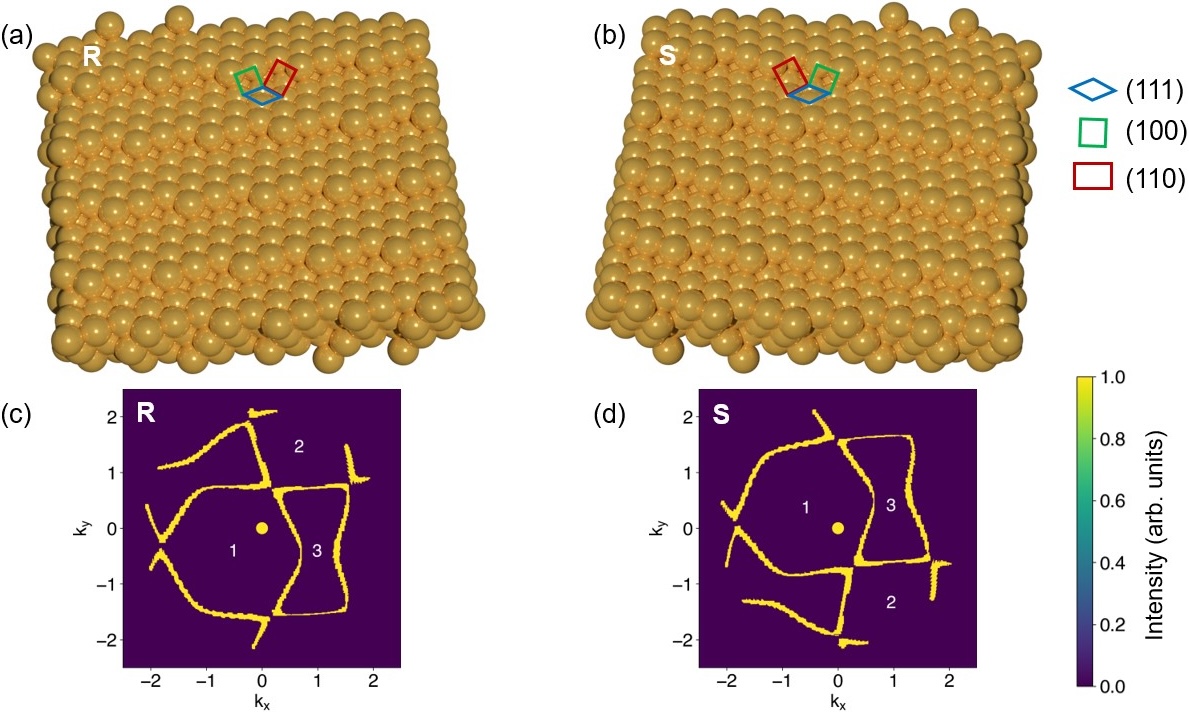}
\caption{Representative images from the two datasets.
(a,~b)~Atomic models of an R and S fcc kink surface.
The three microfacet types at the kink site are indicated by
coloured labels: (111) in blue (priority~1), (100) in green
(priority~2), and (110) in red (priority~3).
Tracing the sequence $1\!\to\!2\!\to\!3$ proceeds clockwise for (a)
the R surface and anticlockwise for (b) the S surface.
(c,~d)~Corresponding simulated ARPES Fermi surface projections for
the same Miller index pair (simulated for h$\nu=21.2\,$eV). White numerals 1, 2, and 3 mark the
Fermi surface features associated with the (111), (100), and (110)
microfacets, respectively. The mirror symmetry breaking
between (c) and (d) is evident in the reversal of the rotational
sense of these features about the surface normal projection (filled
circle), which coincides with the Brillouin zone centre $\bar{\Gamma}$
for the surface shown here.}
\label{fig:datasets}
\end{figure}

Chiral handedness at fcc metal surfaces is assigned using the
microfacet notation and the Cahn--Ingold--Prelog priority rules.\cite{crystalnomenclature,
crystalnomenclature2}
Each kink site exposes three distinct microfacet types: (111), (100), and
(110), denoting hexagonal, square and rectangular atomic arrangements, respectively as illustrated in Figure~\ref{fig:datasets}. These are then assigned priorities 1, 2, and 3 in descending order of surface atom density.
Tracing the sequence $1\!\to\!2\!\to\!3$ in the plane of the surface yields
a clockwise sense for R surfaces and an anticlockwise sense for S surfaces,
as shown in Figure~\ref{fig:datasets}(a,b).
% For any R surface $(h\,k\,l)$, the enantiomeric S surface is obtained by
% inversion of all Miller indices to $(\bar{h}\,\bar{k}\,\bar{l})$.

We highlight an analogy that, to our knowledge, has not been explicitly
noted before. The priority sequence described above for real space is also preserved in reciprocal space, where it is possible to visually identify similar hexagonal, square and rectangular features.
These Fermi surface features associated with the three microfacets rotate in
opposite senses about the surface normal projection (filled circle) for the R and S enantiomers, making the handedness directly readable from the momentum
map $(k_x, k_y)$, as shown in Figure~\ref{fig:datasets}(c,d).
The surface normal projection in an ARPES map plays the same anchoring
role as the kink atom position in real space: it defines the reference
frame from which the rotational sense of the Fermi surface polygons is
inferred and the R/S assignment is made.

To train and evaluate ML classifiers on both representations, we constructed
two independent labeled image datasets spanning a range of chiral fcc
Miller index families.
The first dataset consists of real space atomic model images of the different surfaces.
The second dataset consists of simulated ARPES Fermi surface projections
in reciprocal space.
Both datasets carry two-class labels (R and S) and are deliberately small,
reflecting the data-scarce regime of novel surface science experiments.

The two representations are not equivalent in how they are viewed.
The Fermi surface maps are intrinsically two-dimensional momentum space
images. By contrast, the real space atomic models are two-dimensional projections of a three-dimensional surface, so the apparent kink geometry depends largely on the viewing direction. This makes the real space task inherently more sensitive to perspective.

\begin{figure}
\centering
\includegraphics[height=12cm]{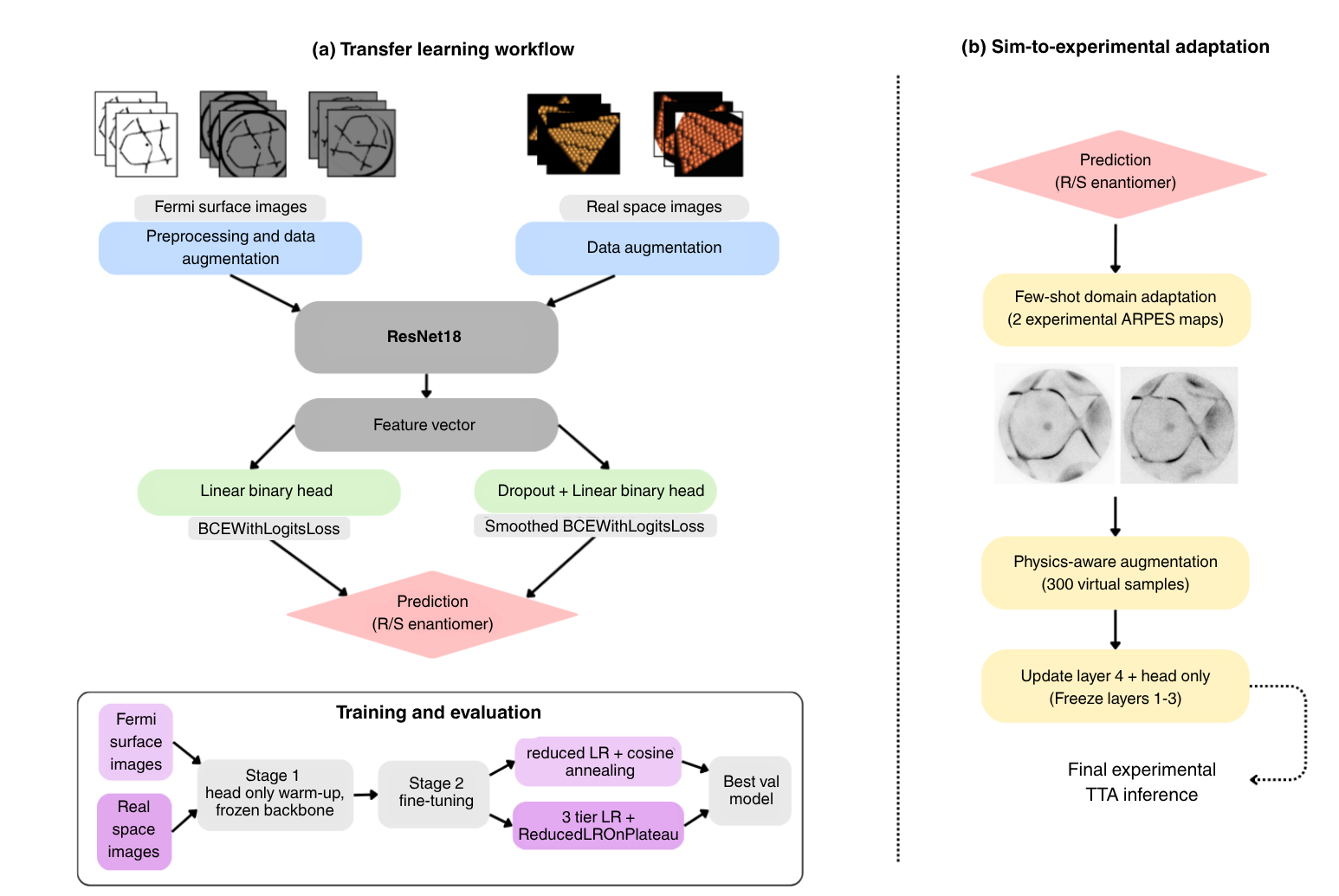}
\caption{Transfer learning strategy for \(R/S\) chirality classification. Workflow for (a) training on synthetic real space and reciprocal space (ARPES) images, and (b) sim-to-experimental adaptation of the pretrained classifier.}
\label{fig:flowchart}
\end{figure}

The overall transfer learning pipeline is illustrated in Figure~\ref{fig:flowchart}, with complete details found in the Methods and Supporting Information. The workflow combines synthetic dataset generation, physics-aware data augmentation, transfer learning,
and experimental validation within a common classification pipeline.
Real space atomic models and reciprocal space Fermi surface projections
are generated independently, split into training, validation, and test datasets, then used to train separate binary R/S
classifiers on a ResNet18 backbone.
The reciprocal space classifier is subsequently applied to experimental
ARPES maps, and is fine-tuned only
on a small labeled dataset.
This sequence captures the full sim-to-experiment (simulation-to-experiment) transfer strategy used
throughout the study.

\subsection{Real Space Classifier}

\begin{figure}[ht]
\centering
\includegraphics[width=\linewidth]{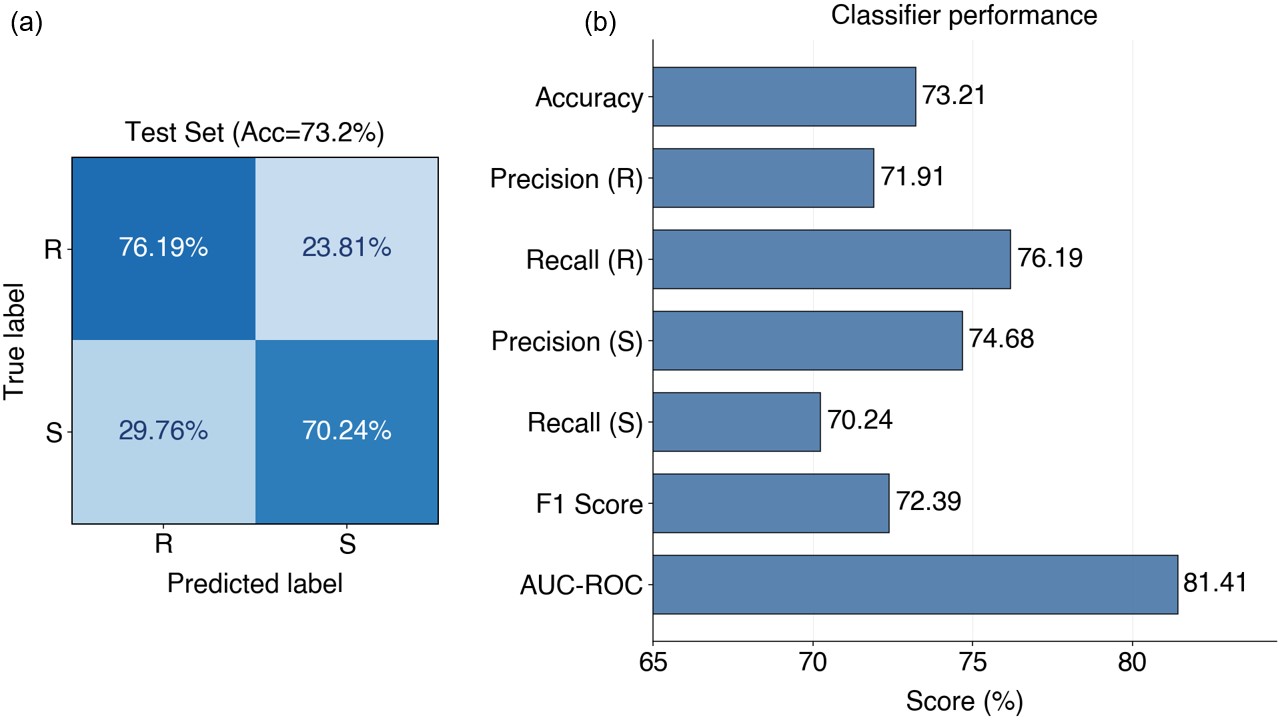}
\caption{Held-out test set performance of the real space \(R/S\) classifier. (a) Normalized confusion matrix showing the fraction of real space images assigned to each predicted class, with diagonal entries corresponding to correct classification of the two enantiomers. (b) Summary performance metrics for the same test set, including accuracy, class-specific precision and recall, F1 score, and AUC--ROC.}
\label{fig:SI1}
\end{figure}

The real space dataset comprises atomic model images of R and S fcc kink
surfaces rendered for Cu, Au, and Pt substrates across 21 Miller index
families.
Image diversity was introduced by independently varying the surface metal,
adatom identity (H, Fe, Pb), adatom coverage, and the presence of surface
defects, yielding a total of 1,662 labeled images across clean surfaces,
adatom-decorated surfaces, defect-containing surfaces, and combinations
thereof.
Each surface was rendered from 15 distinct viewpoints to ensure
comprehensive coverage of the stereocenter geometry.

The chirality signal in real space images is localized at the kink sites,
which occupy only a small fraction of the total image area.
Depending on the viewing angle, the kink geometry can be clearly resolved
or partially obscured, with some perspectives rendering the
step--kink--terrace arrangement ambiguous.
The variation in adatom identity further modifies the apparent
local geometry at the kink site, introducing additional variance that
the classifier must learn to discount.

Beyond the classification challenge, reading surface handedness from
real space measurements is experimentally demanding. 
The closest real space experimental analogue is scanning tunneling
microscopy.
The step--kink--terrace
arrangement is sensitive to thermal roughening, which reduces kink density
and blurs the local chiral motif at realistic surface
temperatures.\cite{realstructure643, gellman2021}
These factors make unambiguous handedness assignment from real space
data alone a persistent experimental challenge.

ResNet18, pretrained on ImageNet-1k, was fine-tuned on this model dataset
using a two-stage training strategy with physics-aware augmentation
that explicitly accounts for the chirality-inverting effect of mirror
operations (see Methods). The classifier's performance were evaluated using the two-class confusion matrices and quantitative
performance metrics as shown in Figure~\ref{fig:SI1}. The confusion matrix is shown in Figure~\ref{fig:SI1}(a), where the diagonal entries denote correct predictions and the off-diagonal entries denote misclassifications. The classifier
accuracy, which measures how often the model makes a correct prediction, is 73.2\,\%  on the held-out test set. The precision, which is the proportion of all the model's positive classifications that are actually positive is 73.6\,\% for R and 72.7\,\% for S. The recall, which is the proportion of all actual positives that were classified correctly as positives, is 72.0\,\% for R and 74.3\,\% for S. The
macro F1 score, which evaluates the model’s average performance across classes, is 73.1\,\%. The AUC-ROC (Area Under the Receiver Operating Characteristic) for binary classification problems, representing the degree of separability between classes, is  0.81 (Figure~\ref{fig:SI1}(b)) \cite{google_mlcc}. These numbers are meaningful and well above chance, confirming that the classifier
has learned a genuine chirality signature from the atomic geometry.
However, it falls short of definitive assignment which is physically
grounded in the spatial locality and viewing-angle sensitivity of the
kink-site chirality cue.
This motivates a complementary approach in reciprocal space, where the
chirality signature is distributed two-dimensionally across the entire image.

\subsection{Fermi Surface Classifier}

% \textcolor{red!60!black}{Early identification of chiral surfaces relied on
% two-dimensional diffraction patterns such as low-energy electron diffraction (LEED)\cite{crystalnomenclature}. This inspires the question of whether the R/S classification convention can be
% generalized across representations. ARPES provides a natural momentum space counterpart, because it maps the
% electronic structure into a two-dimensional reciprocal space image that
% can be compared directly with the synthetic Fermi surface projections.}

\begin{figure*}[t]
    \centering
    \includegraphics[width=\textwidth]{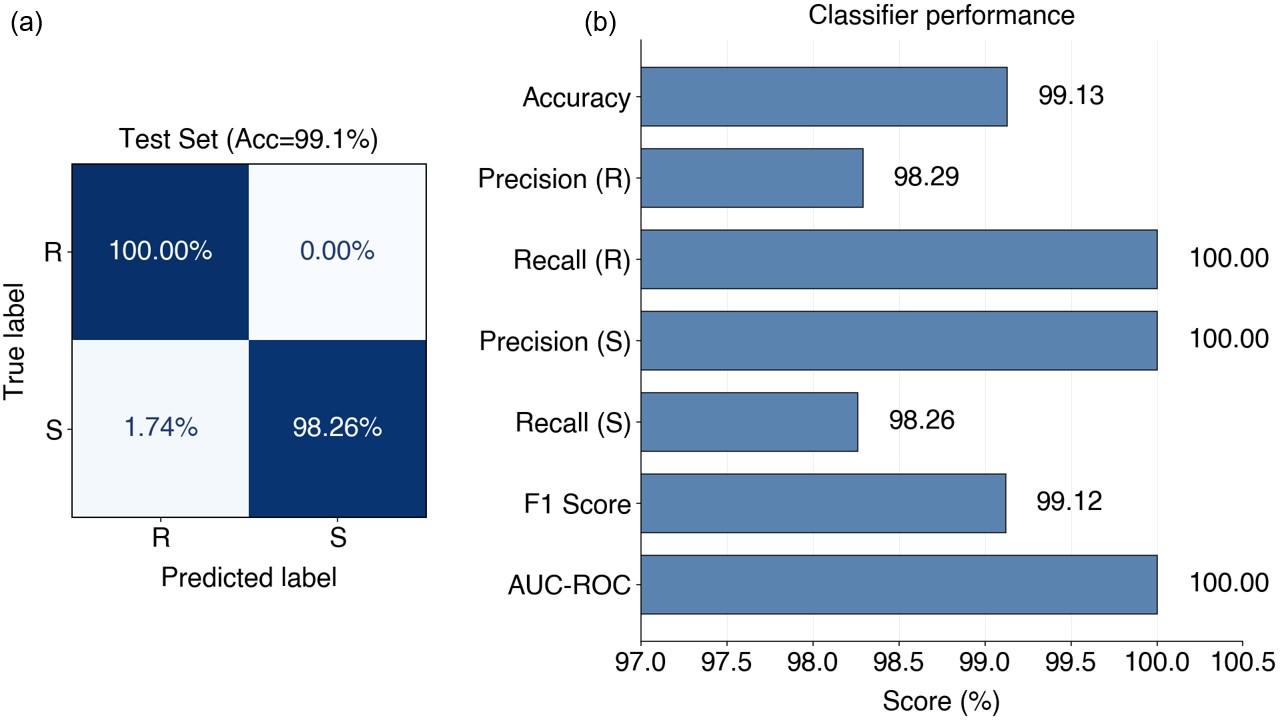}
    \caption{Held-out test set performance of the reciprocal space \(R/S\) classifier. (a) Normalized confusion matrix showing the fraction of Fermi surface images assigned to each predicted class, with diagonal entries corresponding to correct classification of the two enantiomers. (b) Summary performance metrics for the same test set, including accuracy, class-specific precision and recall, F1 score, and AUC--ROC.}
    \label{fig:fermiworkflowmetrics}
\end{figure*}

The reciprocal space dataset comprises simulated ARPES Fermi surface
projections for Cu surfaces across 26 Miller index families, covering
both R and S enantiomers for each family.
Fermi surface cuts at a photon energy of 21.2\,eV were simulated by
intersecting a precomputed three-dimensional Fermi surface mesh with
the free-electron photoemission detection sphere, using the
free-electron final-state approximation for the perpendicular
momentum.\cite{hufnerARPES, ARPES_Springer_2015, ARPES2019}
The dataset contains 3,432 labeled images in total. The details of the
simulation parameters are given in the Supporting
Information. Image diversity was
introduced by sampling a broad set of chiral $(h\,k\,l)$ surface families with varying parameters and
generating both R and S enantiomers for each family, so that the classifier
learned the handedness signature across a range of reciprocal space
geometries rather than from a single surface type.

Experimental ARPES data on intrinsically chiral metal surfaces are
rare, making it impractical to build a training set from measured
maps alone.
Simulated Fermi surface projections bridge this gap because the
chirality signature in reciprocal space arises from the global
orientation of Fermi surface polygons relative to the surface normal
projection. It is fully captured by the simulation without requiring
absolute photoemission intensities. Before applying the classifer to experimental data, we check the validity of the classifier with the synthetic data, suitably augmented and processed.
% A deterministic preprocessing pipeline was applied to all images
% to standardize the input representation: each image was converted
% to grayscale, masked with a circular aperture, binarized using
% adaptive Gaussian thresholding, and thinned to single-pixel-wide
% arc contours (see Methods).
% This binarization eliminates domain-specific intensity variations
% between simulated and experimental images, providing a natural form
% of domain normalization without any explicit domain-adaptation step
% during inference.

ResNet18, pretrained on ImageNet-1k, was fine-tuned on the simulated
Fermi surface dataset using the same physics-aware augmentation logic as the real space classifier and additional pre-processing to enable sim-to-experiment transfer (see Methods).
The classifier for the synthetic Fermi surface data achieves 99.1\,\% accuracy on the held-out test set,
with a macro F1 score of 99.1\,\%, precision of 98.3\,\% for R and
100.0\,\% for S, recall of 100.0\,\% for R and 98.3\,\% for S, and
an AUC--ROC of 1.00. This is shown, along with the two-class confusion matrix in Figure~\ref{fig:fermiworkflowmetrics}.
This near-perfect performance reflects a fundamental difference in
how chirality is encoded in the two representations.
In real space, the chirality cue is local, confined to the kink site, and
is sensitive to viewing angle, adatom identity, and local structural
disorder.
In reciprocal space, mirror symmetry is lifted cleanly and globally
across the entire Fermi surface image, with features associated with
all three microfacets simultaneously carrying the handedness signature in the two-dimensional image.
The reciprocal space representation is therefore intrinsically less
sensitive to local noise.
This resilience has a direct physical parallel in the behavior of
geometric CISS under disorder: spin selectivity persists under strong
Anderson disorder in solid-state chiral systems because the chiral
signal reflects the collective boundary geometry rather than any
individual scattering site.\cite{2026_CISSdevicePRBL}
In the same way, thermal roughening at realistic metal surfaces reduces
kink density while leaving the net chirality of the surface intact,
and ARPES captures this net chirality globally across the full surface
Brillouin zone rather than sampling individual kink
sites.\cite{realstructure643}
The handedness signature is therefore encoded collectively and
redundantly across the entire Fermi surface map, providing the
physical basis for the robustness of the reciprocal space classifier.

%%%%%%%%%%%%%%%%%%%%%%%%%%%%%%%%%%%%%%%%%%%%%%%%%%%%%%%%%%%%%%%%%%%%%
%\section{Results and Discussion}
%%%%%%%%%%%%%%%%%%%%%%%%%%%%%%%%%%%%%%%%%%%%%%%%%%%%%%%%%%%%%%%%%%%%%

\subsection{Experimental Validation}

\begin{figure}
\centering
%% PLACEHOLDER: Figure 3
%% Two panels:
%% (a) Experimental ARPES map for the R surface, annotated with
%%     classifier-predicted label ("R") and sigmoid confidence score.
%% (b) Experimental ARPES map for the S surface, annotated with
%%     classifier-predicted label ("S") and sigmoid confidence score.
%% Each panel also shows the surface normal marker (filled circle)
%%     used as the geometric reference for the classifier.
\includegraphics[width=\textwidth]{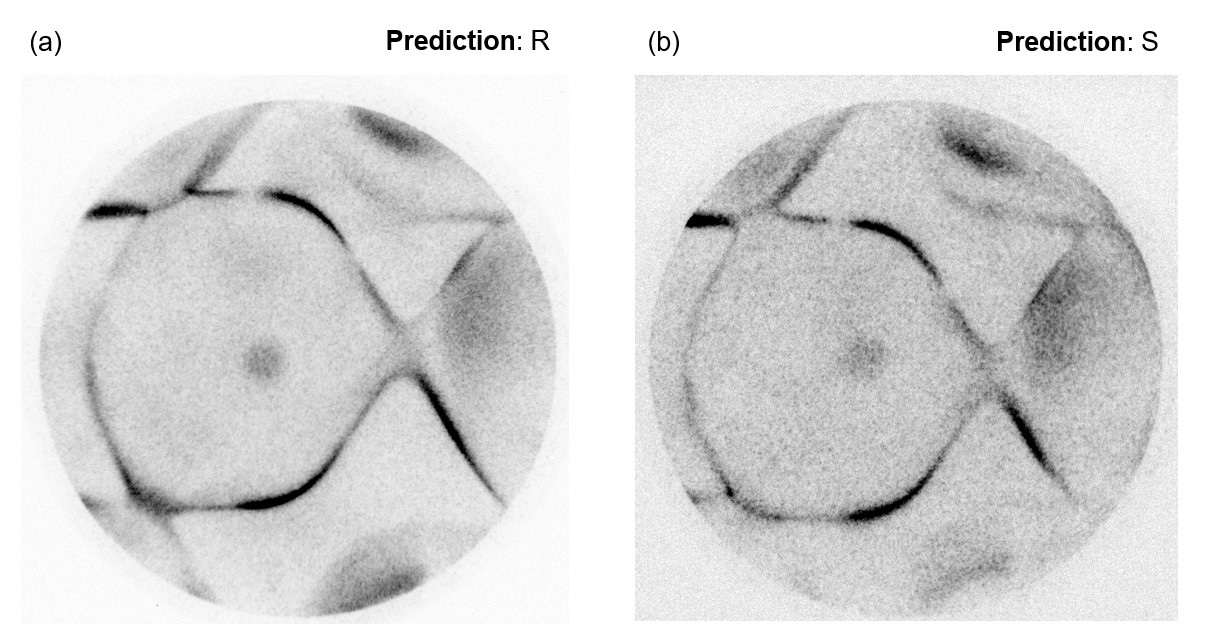}
\caption{Experimentally acquired Fermi surface maps for the (a) R and (b) S
surfaces of Cu(643), respectively (h$\nu=30\,$eV). Each panel is annotated with the
classifier-predicted label.}
\label{fig:arpes_exp}
\end{figure}

To assess real-world applicability, the trained Fermi surface classifier
was applied to two experimental ARPES Fermi surface maps acquired for
Cu(643)$^R$ and Cu(643)$^S$. Fermi surface maps of this kind can be acquired at room temperature, making them experimentally accessible without the need
for low-temperature measurements..
These maps were acquired at a photon energy of 30\,eV, distinct from
the 21.2\,eV used in the simulated training set.
This mismatch changes the absolute photoemission intensities and the
detection geometry, yet the classifier transferred successfully,
confirming that it learned the topological chirality pattern rather than
photon-energy-specific intensity features.

A key challenge in transferring from simulated to experimental images
is the difference in contrast, noise characteristics, and spectral
sharpness between the two domains.
To bridge this gap, both the simulated training images and the
experimental maps were subjected to the same deterministic
preprocessing pipeline, reducing both to binarized arc contours on a
uniform background.
In experimental images, the surface normal projection is identified by
the surface state appearing at $\bar{\Gamma}$, which serves the same
geometric anchoring role as the embedded marker in the synthetic
training images.

In a zero-shot forward pass, the synthetic-trained model correctly
classified both experimental images, reporting 79.8\,\% confidence
for the R enantiomer and 84.3\,\% for the S enantiomer.
To improve prediction confidence, the model was lightly fine-tuned on
augmented copies of the two labeled experimental images, with only the
final residual block and classification head updated to prevent
catastrophic forgetting of synthetic-domain knowledge (see Methods).
After fine-tuning, confidence increased to 99.7\,\% for the R
enantiomer and 100.0\,\% for the S enantiomer.
To confirm that these predictions are stable across all azimuthal
orientations and are not an artefact of a single favourable image
alignment, rotation-averaged inference was performed over 30 evenly
spaced in-plane angles.
This yielded a confidence of 99.2\,\%\,$\pm$\,1.3\,\% for the R enantiomer and
98.9\,\%\,$\pm$\,2.7\,\% for the S enantiomer, with no
misclassification at any tested rotation angle
(Table~\ref{tab:exp}, Figure~\ref{fig:arpes_exp}).

The successful application of a classifier trained on simulated data to synchrotron-acquired experimental data in a single pass establishes that the classifier has learned to recognise and classify handedness for high-Miller index surfaces. This demonstrates ARPES as a quantitative chirality readout for
intrinsically chiral metal surfaces, without requiring atomic-resolution
real space imaging.

\begin{table}
\centering
\caption{Experimental ARPES image classification results.
Single-pass and rotation-averaged TTA confidence are reported for the synthetic-trained model and the
lightly fine-tuned model.
TTA was applied only to the fine-tuned model.}
\label{tab:exp}
\begin{tabular}{llccc}
\toprule
Model & Image & Prediction & Single-Pass & TTA ($n = 30$) \\
\midrule
Synthetic-trained & R enantiomer & R~\checkmark & 79.8\,\% & --- \\
Synthetic-trained & S enantiomer & S~\checkmark & 84.3\,\% & --- \\
Fine-tuned        & R enantiomer & R~\checkmark & 99.7\,\% & 99.2\,\%\,$\pm$\,1.3\,\% \\
Fine-tuned        & S enantiomer & S~\checkmark & 100.0\,\% & 98.9\,\%\,$\pm$\,2.7\,\% \\
\bottomrule
\end{tabular}
\end{table}

Relatedly, our real space comparison used atomic structure models rather than experimental STM data. This is a distinction that matters as experimental STM signals convolve the local density of states with tip geometry, so even genuine topographs do not directly measure atomic-scale surface geometry. Disentangling electronic contrast from true geometric structure is required to extract handedness unambiguously from such data. This additional interpretive step is not required for Fermi surface maps because they encode handedness directly through momentum space polygon topology. This explains why a classifier trained purely on simulated data generalizes so effectively to experimental ARPES spectra and suggests that momentum space may be better suited to this class of chirality-recognition problems.

\section{Conclusion}
We have shown that crystallographic surface chirality in intrinsically
chiral fcc metals can be decoded from both real space atomic models and simulated ARPES Fermi surface projections.
A transfer learning classifier trained on each representation assigns
the R/S label successfully for both. The reciprocal space model outperforms the real space model as it encodes chirality across the entire Fermi surface map. Significantly, it also successfully transfers to experimentally acquired ARPES data. Our core result is that handedness is preserved in momentum space and can be decoded using electronic structure alone. The global encoding of geometric surface chirality in electronic momentum space enables the ML classifier to unambiguously differentiate the enantiomers.

% The approach has minimal data requirements, since both models use a
% purely vision-pretrained backbone with no chemical prior and generalize
% across a broad set of Miller index families.

Several extensions follow naturally.
One promising direction is circular dichroism in the angular distribution
(CDAD) of photoelectrons from matched R and S surfaces.
Such measurements could test whether the handedness encoded in the Fermi
surface topology is also reflected in the chiral photoemission response.
They may also clarify whether classifier confidence tracks the strength
of the underlying chiral electronic signature.
If so, the present framework could provide a practical route to screen
chiral metal surfaces for CISS-related behavior without requiring
spin-resolved measurements.
More broadly, the same workflow should extend to other fcc metals and,
after retraining, to bcc surfaces, while future datasets that include
adsorbates, molecular overlayers, or oxide layers could broaden the
approach toward more realistic catalytic and spintronic interfaces.

%%%%%%%%%%%%%%%%%%%%%%%%%%%%%%%%%%%%%%%%%%%%%%%%%%%%%%%%%%%%%%%%%%%%%
\section*{Methods}
%%%%%%%%%%%%%%%%%%%%%%%%%%%%%%%%%%%%%%%%%%%%%%%%%%%%%%%%%%%%%%%%%%%%%

\paragraph{Real Space Image Generation:} Atomic model images were constructed in
Avogadro\cite{avogadro} and Blender\cite{blender} softwares, with image diversity introduced by independently varying the polar, tilt, and azimuthal viewing angles, defects and the surface adatom types. 

%Full details of the rendering pipeline and parameter ranges
%are given in the Supporting Information.

\paragraph{Reciprocal Space Image Generation:} Simulated ARPES Fermi surface projections were generated from an openly available database of three-dimensional Fermi surfaces in VRML format covering 45 elemental solids~\cite{3DFermiSurfaceDatabase}, ensuring reproducibility and generalizability beyond a single material system. 

% The tilt augmentation across a $\pm 2^{\circ}$ grid of
% $(\theta_x, \theta_y)$ combinations, combined with the repositioned
% surface normal marker in each tilted image, taught the classifier to
% locate the nearest microfacet directions relative to the true surface
% normal rather than relative to the fixed detector frame.
% Full details of the projection algorithm, tilt grid, and parameter
% ranges are given in the Supporting Information.

\paragraph{Data splitting:} Both datasets were split into training, validation, and test sets using a 75-15-10 ratio
drawn from the same pool of Miller index families, with test images held out at the image level to provide an unbiased estimate of classification accuracy.

%% NEW SUBSECTION (from manuscript_revised.tex)
\paragraph{Physics-Aware Data Augmentation:}
A key symmetry constraint of crystallographic chirality guided the
augmentation strategy for both classifiers.
A single mirror operation (horizontal or vertical flip) is an improper
rotation that inverts handedness and therefore inverts the class label
($R \leftrightarrow S$), whereas a double mirror (equivalent to a
$180^{\circ}$ proper rotation) preserves chirality and leaves the
label unchanged.
Accordingly, every horizontal and vertical flip applied during
training was accompanied by a label flip, preventing the network from
associating mirror related image pairs with the same class. The complete pre-processing and data augmentation pipeline for each classifier is given in the Supporting Information.

%% UPDATED SUBSECTION — corrected loss/optimizer/LRs from manuscript_revised.tex
\paragraph{Model Architecture and Training:}

ResNet18 pretrained on ImageNet-1k was fine-tuned separately on the
real space and reciprocal space datasets, with training proceeding in two stages for each classifier. In Stage 1, the back-
bone was frozen and only the classification head was trained. In Stage 2, the full network was fine-tuned with differential learning rates, with early stopping invoked for the respective classifiers to prevent overfitting.

\paragraph{Experimental ARPES Data:}

The experimental data for the Cu(643)$^{R}$ and Cu(643)$^{S}$ surfaces were acquired at Elettra synchrotron (Trieste, Italy) at room temperature, at a photon energy of 30 eV. The surfaces were prepared by repeated cycles of argon ion sputtering and sample annealing at 500 $^\circ$C.

\paragraph{Zero-shot classification:}
The synthetic-trained model was applied directly to the two
preprocessed experimental images without any retraining or modification of model weights, yielding
correct classification in a single forward pass for both enantiomers
(Table~\ref{tab:exp}).

%% UPDATED PARAGRAPH — combined detail from both files
\paragraph{Few-shot fine-tuning.}
To improve prediction confidence, the model was fine-tuned on
augmented copies of the two labeled experimental images. Rotation averaging tests whether the classification is stable across
all in-plane orientations, which is expected from a physically
meaningful chirality classifier that is not biased toward any
particular azimuthal alignment.

% \subsection*{Data Availability}

% \textcolor{blue!60!black}{All image datasets and trained model weights are deposited under a CC-BY 4.0 license on Zenodo at
% \url{https://doi.org/10.5281/zenodo.21923938}. The inference scripts along with the complete code is available under a GPL 3.0
% license on Github at \url{<link to github repo>}. A MaRDI Packaging System (MaPS) [\url{https://arxiv.org/abs/2404.05563}][\url{https://maps.mardi4nfdi.de/}]
% runtime is published as \url{io.shtekur.classifierfermi/x86_64/v1}. The runtime enables a full reproduction of all results
% on a standard Linux system without the need of manual dependency resolution.}

%%%%%%%%%%%%%%%%%%%%%%%%%%%%%%%%%%%%%%%%%%%%%%%%%%%%%%%%%%%%%%%%%%%%%
\begin{acknowledgement}
The authors thank Vitaliy Feyer and Iulia Cojocariu
 for assistance with synchrotron data acquisition, M. S. Santhanam and Nisarg Vyas for feedback on the manuscript, and Anirudh B V for infrastructure support.
 The experimental maps were supported by the Deutsche Forschungsgemeinschaft (DFG, German Research Foundation) ---TRR 173-268565370 Spin + X: spin in its collective environment (Project B14). A.K. acknowledges support by MaRDI, funded by the Deutsche Forschungsgemeinschaft (DFG), project number 46015501, NFDI 29/1 "MaRDI - Mathematische Forschungsdateninitiative".
\end{acknowledgement}

\begin{suppinfo}
Table of Miller indices and handedness assignments; Real and reciprocal space image rendering pipeline and parameter ranges;
Method for simulation of Fermi surface cuts; Full transfer learning pipeline with pre-processing and augmentation strategies, and all hyperparameter settings.
\end{suppinfo}

\bibliography{achemso-demo}

\includepdf[pages=-]{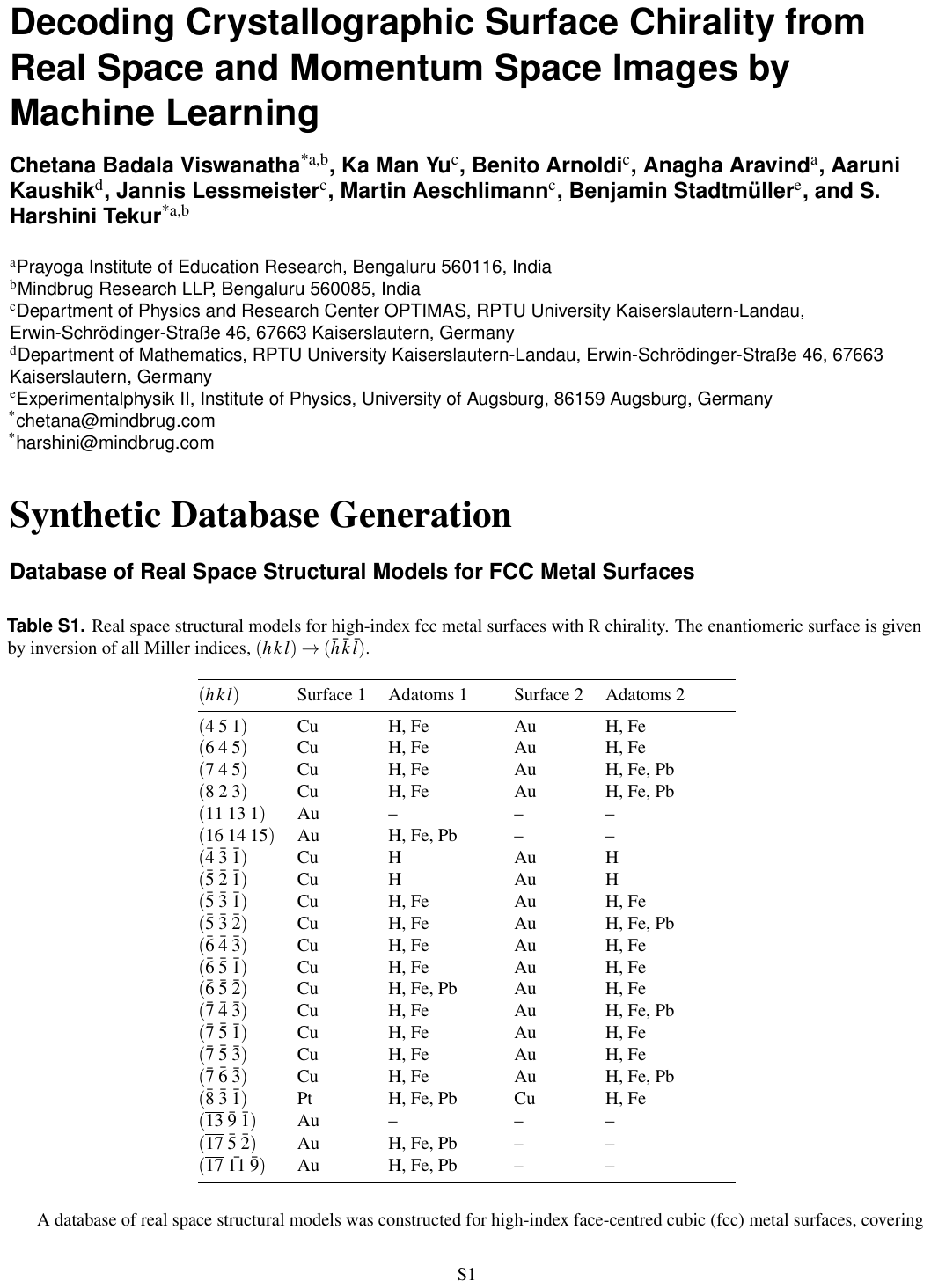}

\end{document}